\documentstyle[prl,aps,epsf]{revtex}
\begin{document}
\draft
\title{Delocalized Quasiparticles in the Vortex State of an Overdoped High-$T_c$ Superconductor Probed by $^{63}$Cu NMR }

\author{G.~-q.~Zheng$^1$, H.~Ozaki$^1$, Y.~Kitaoka$^1$, P. Kuhns$^2$, A. P. Reyes$^2$, and W. G. Moulton$^2$}
\address{$^1$Department of Physical Science, Graduate School of Engineering
Science, Osaka University, Toyonaka, Osaka 560-8531,
Japan}
\address{$^2$National High Magnetic Field Laboratory, Tallahassee, FL
32310}

\twocolumn[
\maketitle
\widetext
\begin{abstract}
We report the spin Knight shift ($K_s$) and the nuclear spin-lattice relaxation rate ($1/T_1$) in the vortex state as a function of magnetic field ($H$) up to 28 T in the high-$T_c$ superconductor TlSr$_{2}$CaCu$_2$O$_{6.8}$ ($T_c$=68 K). At low temperatures well below $T_{c}$, both $K_s$ and $1/T_1$ measured around the middle point between two nearest vortices (saddle point) increase substantially with increasing field, which indicate that the quasiparticle states with an ungapped spectrum are extended outside the vortex cores in a d-wave superconductor. The density of states (DOS) around the saddle point is found to be $\kappa N_0\sqrt{H/H_{c2}}$, with $\kappa$=0.5$\sim$0.7 and $N_0$ being the normal-state DOS.
\end{abstract}
\vspace*{5mm}
]
\narrowtext

The vortex state in unconventional superconductors, such as the high transition-temperature (high-$T_c$) superconducting  copper oxides which have a $d_{x^2-y^2}$-wave gap symmetry \cite{review}, is of great interest.
Among various issues,  the quasiparticle state
associated with the vortices still remains far from being well understood.  In an s-wave superconductor with isotropic gap, the low-lying quasiparticle states are localized within the vortex core whose radius is $\xi=v_{F}/\pi \Delta_{0}$, with energies $E=\mu\Delta_{0}^{2}/E_{F}$  ( $\mu$=1/2,3/2....) \cite{Caroli}, where $\Delta_{0}$
is the energy gap, $v_{F}$ and $E_{F}$ are the velocity and the energy level
at the Fermi surface, respectively. This prediction was verified experimentally by the scanning tunneling microscopy (STM) \cite{Hess}.   In contrast, the gap for d-wave superconductivity
vanishes in the nodes which leads to a divergent $\xi$ along the nodal directions. The corresponding quasiparticle state is, therefore, expected to have very different properties.
Recent theories suggest that the quasiparticle states may be  extended far outside the
cores along the nodal directions \cite{Volovik,Wang,Won,Franz,Ichioka}. However, this prediction has not yet been unambiguously confirmed by experiments. Heat capacity measurement at low fields in the vortex state of YBa$_2$Cu$_3$O$_7$ (YBCO) has revealed a $\sqrt{H}$ dependence of the specific heat coefficient \cite{Moler}, which was taken as evidence for supporting the theoretical prediction \cite{Volovik}. But heat capacity also contains contributions from the vortex cores which may have a $\sqrt{H}$-like variation in the low field regime as observed in s-wave superconductors and pointed out by Sonier {\em et al} \cite{Sonier}. Our earlier NMR measurement on an underdoped cuprate \cite{Zheng} and thermal transport measurement \cite{Chiao} in the  low $H/H_{c2}$ regimes are consistent with the existence of finite density of quasiparticle states outside the cores, but  no quantitative estimation of the field-induced states has been available. The most serious question about the quasiparticle picture in a d-wave superconductor is raised recently by  STM experiments that find no quasiparticle states outside the cores \cite{Maggio,Pan}. Clearly, more spatially-resolved and bulk-sensitive studies, and also at higher fields,  
 are needed to resolve this issue.

We report in this Letter measurements of the spin susceptibility and the nuclear spin-lattice relaxation rate ($1/T_1$) outside the vortex cores as a function of $H$ up to $H$=28 T (or $H/H_{c2}$=0.65) in a d-wave high-$T_{c}$ cuprate. We find compelling evidence for the quasiparticles in the vortex state being delocalized. Unlike  specific heat measurement that picks up contributions from quasiparticles both inside and outside the vortex cores, NMR technique can measure separately the quasiparticle states  in- and out-side the vortex core.  In the mixed state,  those nuclei located in between two nearest vortex cores have the largest distribution probability that gives rise to a singularity in the field distribution function, while those in the cores  contribute to a "knee" that is far away from the singularity \cite{Maclaughlin}.   In practice, however, both the singularity and the knee are usually smeared out owing to, {\em e.g. } lattice disorders,  resulting in a more symmetric NMR line shape. Nevertheless, some information about the position selectiveness is retained; the peak of the NMR line that results from  smeared singularity corresponds to the nuclei located outside the vortex cores and in the superconducting state.

The sample chosen for this study is a slightly overdoped cuprate TlSr$_{2}$CaCu$_2$O$_{6.8}$ ($T_c$=68 K). This compound has several advantages over the most-extensively-studied YBCO family with regards to our purpose. First, the  Knight shift arising from the spin susceptibility  when field is applied parallel to c-axis is large \cite{Zheng1}, which allows an accurate measurement of its change in the superconducting state in the presence of  magnetic fields. This is to be compared to the YBCO family where the corresponding shift is negligibly small  because of the accidental cancellation of the hyperfine field, so that measurement along this direction is not possible \cite{Zheng}. Second, the spin susceptibility in the normal state above $T_c$ is almost temperature independent, which helps to evaluate the density of states (DOS) in the normal state. Third, the upper critical field $H_{c2}\sim$ 43 T is relatively small \cite{Zheng1}, so that the field-induced DOS at a given $H$, if any, should be larger. Finally, there is only one crystallographic site of Cu, which greatly simplifies NMR spectroscopy and enhances data accuracy.

The polycrystals were c-axis aligned under a field of 16 T and fixed with a polymer. For all measurements, the external field is applied along the c-axis. Figure 1 shows a typical NMR line shape at $T$=4.2 K (gray curve). The dotted curve is the corresponding theoretical field distribution for 75$^{\circ}$ vortex lattice (VL) obtained as described below. The position-dependent field $H(r)$ in the sample was calculated by using the London model \cite{Brandt},
\begin{eqnarray}
H(r) & = &\bar{H} \sum_{l,m} \frac{ \exp(-G_{lm}^2 \xi^2/2) \exp(-i\vec{G_{lm}} \cdot \vec{r}) }{ 1+G_{lm}^2\lambda^2 }
\end{eqnarray}
where $\bar{H}$ is the averaged  (externally applied) field, $\xi$ and $\lambda$ are the coherence length and penetration depth, respectively. The summation runs over all reciprocal VLs  $\vec{G_{lm}}=2\pi\sqrt{\frac{\bar{H} Sin \beta}{\phi_{0}} } (m\hat{x}+\frac{1}{ Sin \beta }(l-m \cdot Cos \beta ) \hat{y})$, where $\beta$ is the angle between two primitive VL vectors and $\hat{x}$ and $\hat{y}$ are the unit vectors of the reciprocal VL. We adopt $\beta=75^{\circ}=5\pi /12$, which is the value found in neutron scattering \cite{Keimer} and STM \cite{Maggio} experiments. The NMR spectrum is the spatial average of a Dirac delta function,
$f(H') =\langle \delta(H'-H(r)) \rangle _{r}$.
By use of the relation of $H_{c2}=\phi_{0}/2\pi\xi^2$,
eq. (1) is left with a {\em single parameter}, $\lambda$, that is to be determined by experiment. In the theoretical curve of Fig. 1 (dotted curve) calculated from eq. (1), the singularity $S$ corresponds to the middle point between two nearest vortices (saddle point), $M$ and $m$ correspond to the  vortex core and the middle point between four vortices, respectively.
The solid thick curve is a calculation by convoluting eq. (1) with a Lorentzian broadening function 
$p(H')=\frac{\sigma^2}{\sigma^2+4\pi^2 H'^2}$, with a fixed value of $\sigma/\pi$=170 Oe that is the full width at half maximum (FWHM) of the NMR spectrum at $T$=70 K (above $T_{c}$).  In performing the fitting, both the experimental spectrum and the theoretical field distribution curve are normalized to unity. From the peak position, we determined the Knight shift, $K_c$, around the saddle point. We also obtained $\lambda$=950 $\pm$ 100$\AA$ at $T$=4.2 K, and hence the diamagnetic field $H_{dia}=H_{saddle}-\bar{H}$ that depends on the magnitude of externally applied field. Similar value of $\lambda$ was obtained from the analysis of $^{203}$Tl lineshape which has a narrower width, with FWHM = 50 Oe at $T$=70 K and 140 Oe at $T$=4.2 K, respectively.

The value of $\lambda$=950$\pm$100 $\AA$ is slightly smaller than that estimated for the YBCO family (1200$\sim$1400 $\AA$)  \cite{Basov}. This is consistent with the current compound being overdoped. Also, we have  confirmed that the spatially dependent information is indeed retained, even though the singularity and the knee are smeared out.  This was done 
by measuring $1/T_1$ in different positions. $1/T_1$ is larger by a factor of two in the higher end of the spectrum shown in Fig. 1 around the knee, than that measured at the peak.

Now we turn to the main result, the $H$ dependence of the Knight shift and $1/T_1$ measured at the spectrum peak. Figure 2 shows the typical data sets for the $T$ and $H$ dependencies of $K_c$. In the normal state  $K_c$ shows very weak $T$-dependence and does not depend on $H$ above $T$=85 K. In particular, $K_c$ is constant (=1.42$\%$ ) in the temperature range of 85K $\leq T \leq$ 150 K. In contrast, at low temperatures below $T_c$, the shift  strongly depends on $H$. This is clearly shown in Fig. 3 where data at $T$=4.2 K are plotted. The solid circles show the data corrected by the diamagnetic shift $H_{dia}/\bar{H}$ obtained  as described earlier. It can be seen clearly that $K_c-K_{dia}$ increases with increasing field. This corrected shift is composed of the spin part ($K_s$) and the part due to orbital susceptibility $K_{orb}$, respectively, the latter being $H$-independent in the range of field studied. Note that $K_s$ is related to the spin susceptibility and hence the DOS, $N(E)$ as  
$K_s  \propto \int N(E)(-\frac{\partial f(E)}{\partial E})dE$,
where $f(E)$ is the Fermi distribution function. In particular, $K_s$ at low $T$ is proportional to the DOS at the Fermi level, $N(0)$.
Thus, our result indicates that magnetic field induces a finite $N(0)$,  in the superconducting state.

This conclusion is corroborated by the $H$-induced increase of $1/T_1$ measured at the spectrum peak. Figure 4 shows the typical data sets of $1/T_1$ under various fields from $H$=0 to 28 T. The measurement was done by using a single saturation pulse \cite{Zheng} and at the NQR spectrum peak of 22.5 MHz for $H$=0. As seen in the figure,  $1/T_1$ becomes proportional to $T$ below $T\sim$ 10 K, and its magnitude increases with increasing $H$. This is seen more clearly in Fig. 5 where $1/T_1T$ at $T$=4.2 K is displayed. In terms of DOS, $T_{1}$ is expressed as $1/T_{1} = \alpha (T) \int   N(E)^2 f(E)(1-f(E)) dE$, where $\alpha (T)$ is the enhancement factor due to antiferromagnetic (AF) correlation. For a d-wave superconductor \cite{Bulut},  $\alpha (T)$ was shown to have a very weak $T$ variation below $T_c$ and $\alpha (0) \sim 0.9\alpha(T_c)$.  Therefore, $1/T_1T \propto N(0)^2$ at low $T$; the $T_1$ result also indicates that $H$ induces a finite DOS outside the vortex cores. 

Next we evaluate more quantitatively the DOS around the saddle point, $\delta N_H$. First, one notices from Fig. 5 that at high fields of $H\geq$5 T, $1/T_1$ is linear in $H$ which suggests that the induced DOS is proportional to $\sqrt{H}$. In fact, in this field regime, a functional of 
\begin{eqnarray}
 \delta N_{H} &= & \kappa N_0\sqrt{H/H_{c2}} 
\end{eqnarray}
can fit both the Knight shift and $1/T_1$ quite well, where $N_0$ is the DOS in the normal state. In Fig. 3, the solid curve is a fit to
$K_c-K_{dia}=K_{0}+K_H \sqrt{H/H_{c2}}$,
with resultant fitting parameters $K_{0}=1.00\pm0.03 \%$  and $K_H=0.32\pm 0.05\%$.  The solid line in Fig. 4 is 
$\frac{1}{T_1T}=0.018+3.175  \frac{H}{H_{c2}}$ [Sec$^{-1}$K$^{-1}$].  On the other hand, $1/T_1T$ remains finite at $H$=0 and rises more slowly with increasing $H$ at low fields below 5 T.  We ascribe the zero-field residual $1/T_1T$ to that coming from the residual DOS, $\delta N_{imp}$, due to scattering by impurity or crystal imperfection, which is estimated as 
$\frac{\delta N_{imp}}{N_0}=\sqrt{\frac{(T_1T)_{Tc}}{0.9(T_1T)_{4.2K}}}$=0.14. 
Then the slower rise of $1/T_1T$ below 5 T can be naturally understood as vortex flux preferentially occupying the impurity sites where a finite DOS has already been produced \cite{Miyake}. In the impurity-dominated regime, a recent calculation \cite{Vekhter} shows that the field-induced DOS has an asymptotic form of $\delta N_{H} \propto  \frac{H}{H_{c2}}ln (\frac{H_0}{H})$ where $H_0$ is a constant  ($\sim\frac{E_f \Delta_0}{900}$K$^{-1}$). Indeed, the data of $H\leq$5 T can be fitted by such functional as seen in Fig. 5, where the resulting dotted curve is
 $\frac{1}{T_1T} =0.229+1.650 (\frac{H}{H_{c2}}ln \frac{270}{H})^2$
 in unit of Sec$^{-1}$K$^{-1}$. The same $\delta N_{H}$ functional fitting to the shift of $H\leq$5 T with $H_0$=270 T predicts, as shown by the dotted curve in Fig. 3, the shift to remain at 1.03\% when $H$=0.  Since this value is a sum of the impurity-induced spin shift and $K_{orb}$, namely, 
$1.03\%=\frac{\delta  N_{imp}}{N_0}(1.42\%-K_{orb})+ K_{orb}$, we obtain $ K_{orb}$=0.96\%  and hence $K_s$=0.46\%, with uncertainties of $\pm$0.03\% (absolute value) \cite{note2}. Then, $\kappa=K_H/K_s$ is estimated to be 0.70$\pm 0.06\%$. On the other hand, from the $T_1$ result and following the same procedure of estimating $\delta N_{imp}$, the prefactor $\kappa$ is estimated to be 0.50$\pm 0.03\%$. This difference of $\kappa$ by 30\% might arise from the choosing of the AF enhancement factor  $\alpha (0) \sim 0.9\alpha(T_c)$ that underestimates
 $\delta N_{H}$ from the $T_1$ result. The analysis of the Knight shift data could also overestimate $\kappa$; if a triangular model is assumed for the VLs, $\kappa$ would be reduced by $\sim$0.1.

To summarize the above discussions, it is important to stress that both the $H$-enhanced Knight shift and $1/T_1$ observed around the saddle point can be consistently accounted for by a  DOS of  
$\delta N_{H} = \kappa N_0 \sqrt{H/H_{c2}}$, 
with a conservative estimate of $\kappa=0.5\sim$ 0.7.  We therefore conclude that the quasiparticle states are extended outside the vortex cores and that they have a continuous excitation spectrum without a gap. This feature is in sharp contrast with that in an s-wave superconductor. Volovik  first pointed out that the spatially averaged DOS associated with a single vortex in a d-wave superconductor is
$\delta N_{H}=\kappa N_0\sqrt{H/H_{c2}}$,
where  $\kappa$ is on the order of unity \cite{Volovik}.  Although no calculation of the DOS outside the vortex cores is available to allow a direct comparison to our results,  the substantially smaller $\kappa$ than unity found in our experiment could be due to the effects  of vortex lattice, its disorder \cite{Vekhter}, or due to induction of second component of the order parameter \cite{Mao}.

In conclusion, from $1/T_1$ and the spin Knight shift measured around the saddle point in a slightly overdoped high-$T_{c}$ cuprate, we find that  the  quasiparticle states are extended outside the vortex cores with ungapped spectrum in a d-wave superconductor, which is in contrast to the localized states within the vortex cores in s-wave superconductors. The density of such quasiparticle states near the saddle point, in the field range of 5 T$<H<$28 T,  follows $\kappa N_{0}\sqrt{H/H_{c2}}$ where $N_{0}$ is the normal-state DOS. The prefactor $\kappa$ is estimated to be in the range of 0.5$\sim$0.7.  Although such states have not been found in the STM experiments, one may be able to reconcile the STM and NMR results by considering the momentum-dependent tunneling matrix in the STM experiments which likely vanishes in the nodal directions \cite{Wu}.

We thank W.G. Clark, Y. Shimakawa and Y. Kubo for collaborations in the early stage of this work.  G.-q.Z  
 thanks in particular A.V. Balatsky, Y. Matsuda, M. Ogata  and S.H. Pan for several useful conversations, and partial support
 by a Grant-in-Aid for
Scientific Research No. 11640350 from MEXT of Japan and by the Visiting Scientist Program of National High Magnetic Field Laboratory  which is supported by NSF
(DMR-9527035) and by the State of Florida. Additional support from COE research grant No. 10CE2004 of MEXT is acknowledged.

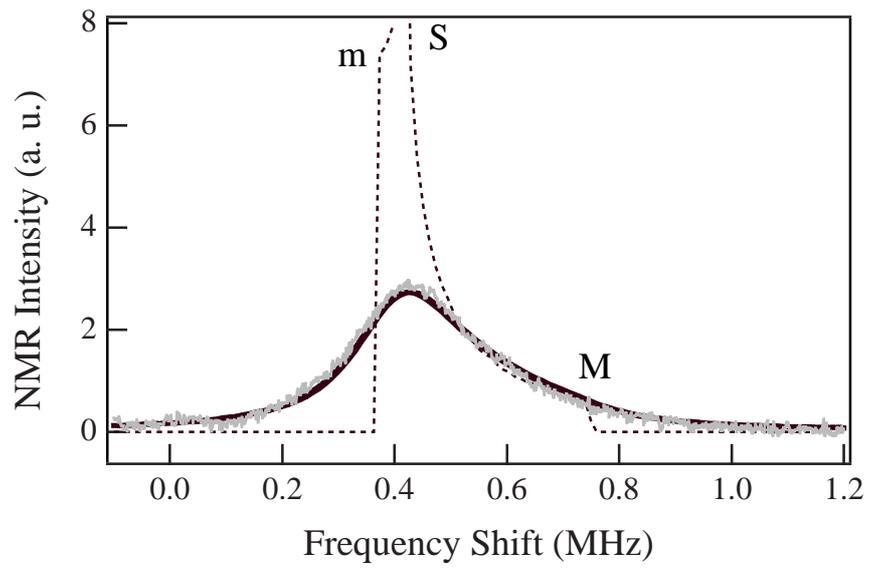
\begin{figure}[htbp]
\caption[]{ $^{63}$Cu NMR line shape at $H$= 4 T in the vortex state (gray curve) where the horizontal axis is measured from $f$=45.14 MHz. The dotted curve is the calculated field distribution for 75$^{\circ}$ vortex lattice. The thick solid curve is a convolution of the dotted one with a Lorentzian broadening function (see text). }
\end{figure}

\begin{figure}[htbp]
\caption[]{Temperature and magnetic field dependencies of the $^{63}$Cu Knight shift. The arrow indicates the superconducting transition temperature at zero field, $T_{c0}$.} 
\end{figure}

\begin{figure}[htbp]
\caption[]{Magnetic field dependence of the $^{63}$Cu Knight shift at $T$=4.2 K. The open and solid circles are the raw data and those after correction for the diamagnetic shift, respectively. The error bars arise from the systematic error in estimating $\lambda$. The solid curve is a fit of the data above 5 T to $K_c(4.2K) -K_{dia}=K_{0}+K_H \sqrt{H/H_{c2}}$. The broken curve is expected when impurity-scattering dominates  (see text). }
\end{figure}

\begin{figure}[htbp]
\caption[]{Temperature and magnetic field dependencies of $1/T_1$. Note that $1/T_1 \propto T$, as indicated by the broken line, below $\sim$10 K at all fields.} 
\end{figure}

\begin{figure}[htbp]
\caption[]{$H$ dependence of $1/T_1T$ at $T$=4.2 K taken from Fig. 4.  The solid line is a linear fit of the data of $H\geq$5 T. The broken curve is $\frac{1}{T_1T} =0.229+1.650 (\frac{H}{H_{c2}}ln \frac{270}{H})^2$ [Sec$^{-1}$K$^{-1}$].}

\end{figure}


\begin{references}
\bibitem{note2}
The $K_{orb}$ is smaller than that in YBCO of 1.28$\%$. This is consistent with the fact that the hole number in the 3$d_{x^2-y^2}$-orbit of this compound is smaller than that in YBCO, as estimated from the NQR frequency \cite {Zheng3}.
\end{references}
\end{document}